\documentclass{l4dc2024}
\usepackage{todonotes}
\usepackage{floatrow}

\title[Hacking Predictors Means Hacking Cars]{Hacking Predictors Means Hacking Cars: Using Sensitivity Analysis to Identify Trajectory Prediction Vulnerabilities for Autonomous Driving Security}
\usepackage{times}

\coltauthor{\Name{Marsalis Gibson} \Email{mtgibson@berkeley.edu}\\
 \Name{David Babazadeh} \Email{davidbabazadeh@berkeley.edu}\\
 \Name{Claire Tomlin} \Email{tomlin@eecs.berkeley.edu}\\
 \Name{Shankar Sastry} \Email{sastry@coe.berkeley.edu}\\
 \addr University of California, Berkeley}

\newcommand{\mgnote}[1]%
    {\textcolor{red}{\textbf{MG: #1}}}

\newcommand{\agentidx}{\ensuremath{i}}
\newcommand{\timeidx}{\ensuremath{t}}
\newcommand{\trajstate}{\ensuremath{s}}
\newcommand{\agentaction}{\ensuremath{u}}
\newcommand{\contextualInfo}{\ensuremath{C}}
\newcommand{\horizon}{\ensuremath{T}}
\newcommand{\predictionInput}{\ensuremath{I}}
\newcommand{\featureidx}{\ensuremath{i}}
\newcommand{\attrfunction}{\ensuremath{f_{\text{attr}}}}
\newcommand{\attrscore}{\ensuremath{c}}
\newcommand{\pertfunc}{\ensuremath{\eta}}

\newcommand{\predstate}{\ensuremath{\hat{\trajstate}}}

\newcommand{\historyhorizon}{\ensuremath{\delta t_h}}

\newcommand{\predictionhorizon}{\ensuremath{\Delta T}}

\newcommand{\dataset}{\ensuremath{\Gamma}}
\newcommand{\datasetidx}{\ensuremath{i}}

\newcommand{\numfeatures}{\ensuremath{{nf}}}
\newcommand{\numpertfuncs}{\ensuremath{{np}}}
\newcommand{\pertfuncidx}{\ensuremath{{i}}}
\newcommand{\datapoint}{\ensuremath{\gamma}}
\newcommand{\resultSet}{\ensuremath{\textit{S}}}
\newcommand{\resultSetidx}{\ensuremath{{i}}}

\newcommand{\qone}{\ensuremath{\text{Q1}}}
\newcommand{\qtwo}{\ensuremath{\text{Q2}}}
\newcommand{\qthree}{\ensuremath{\text{Q3}}}
\newcommand{\agentdynamicsmodel}{\ensuremath{f_{\text{dynamics}}}}

\newcommand{\graph}{\ensuremath{G}}
\newcommand{\nodeSet}{\ensuremath{V}}
\newcommand{\edgeSet}{\ensuremath{E}}
\newcommand{\weightSet}{\ensuremath{w}}
\newcommand{\agentnode}{\ensuremath{A}}

\newcommand{\modelparams}{\ensuremath{\theta}}
\newcommand{\inputtargets}{\ensuremath{y}}
\newcommand{\pertpredictioninput}{\ensuremath{\Tilde{\predictionInput}}}
\newcommand{\perturbation}{\ensuremath{\Tilde{P}}}

\newcommand{\mapfree}{\ensuremath{\mathcal{Z}_{\text{free}}}}
\newcommand{\validHumanDist}{\ensuremath{\epsilon}}
\newcommand{\validplanDist}{\ensuremath{\kappa}}
\newcommand{\numPosStates}{\ensuremath{n_p}}

\newcommand{\predictionFunction}{\ensuremath{f}}
\newcommand{\predictionModel}{\ensuremath{\predictionFunction_{\text{model}}}}
\newcommand{\latentPredVar}{\ensuremath{Y}}
\newcommand{\latentVar}{\ensuremath{Z}}
\newcommand{\latentGiven}{\ensuremath{X}}
\newcommand{\latentDist}{\ensuremath{\psi}}
\newcommand{\latentPredDist}{\ensuremath{\phi}}
\newcommand{\encoder}{\ensuremath{\bar{\predictionFunction}_{encoder}}}
\newcommand{\decoder}{\ensuremath{\Tilde{\predictionFunction}_{decoder}}}

\newcommand{\undetectRes}{\ensuremath{255.50}}
\newcommand{\tplusFGSMEp}{\ensuremath{0.025}}

\newcommand{\currSpeedRes}{\ensuremath{17.6}}
\newcommand{\tplusADEFrom}{\ensuremath{1.34}}
\newcommand{\tplusADETo}{\ensuremath{4.785}}

\newcommand{\AFundetectRes}{\ensuremath{974.0}}
\newcommand{\AFADEFrom}{\ensuremath{0.097}}
\newcommand{\AFADETo}{\ensuremath{1.043}}

\newcommand{\planFGSMep}{\ensuremath{20}}
\newcommand{\planPrevSpeed}{\ensuremath{17.60}}
\newcommand{\planPertSpeed}{\ensuremath{0}}
\newcommand{\planPrevADE}{\ensuremath{1.97}}
\newcommand{\planToADE}{\ensuremath{26.52}}
\newcommand{\planDistNum}{\ensuremath{16.5}}
\newcommand{\humanDistNum}{\ensuremath{15}}
\newcommand{\normperc}{\ensuremath{50}}
\newcommand{\posrange}{\ensuremath{80}}
\newcommand{\velrange}{\ensuremath{30}}
\newcommand{\accelrange}{\ensuremath{35}}
\newcommand{\headrange}{\ensuremath{7}}
\newcommand{\dheadrange}{\ensuremath{5}}
\newcommand{\imagerange}{\ensuremath{1}}
\newcommand{\edgeweightrange}{\ensuremath{10}}

\newcommand{\Tppminordepthres}{\ensuremath{10}}
\newcommand{\AgFminordepthres}{\ensuremath{167}}
\newcommand{\TppUpperQuartileImage}{\ensuremath{553.0}}
\newcommand{\AgFUpperQuartileImage}{\ensuremath{135.0}}
\newcommand{\TppZZOneMedian}{\ensuremath{17.8}}
\newcommand{\AgFZZOneMedian}{\ensuremath{17.3}}
\newcommand{\TppZOneMedian}{\ensuremath{162.0}}
\newcommand{\AgFZOneMedian}{\ensuremath{112.0}}

\newcommand{\TppZZOneMax}{\ensuremath{12800.0}}
\newcommand{\AgFZZOneMax}{\ensuremath{213.0}}

\begin{document}

\maketitle

\begin{abstract}%
 Adversarial attacks on learning-based multi-modal trajectory predictors have already been demonstrated. However, there are still open questions about the effects of perturbations on inputs other than state histories, and how these attacks impact downstream planning and control. In this paper, we conduct a sensitivity analysis on two trajectory prediction models, Trajectron++ and AgentFormer.  The analysis reveals that between all inputs, almost all of the perturbation sensitivities for both models lie only within the most recent position and velocity states. We additionally demonstrate that, despite dominant sensitivity on state history perturbations, an undetectable image map perturbation made with the Fast Gradient Sign Method can induce large prediction error increases in both models, revealing that these trajectory predictors are, in fact, susceptible to image-based attacks. Using an optimization-based planner and example perturbations crafted from sensitivity results, we show how these attacks can cause a vehicle to come to a sudden stop from moderate driving speeds. 
\end{abstract}

\begin{keywords}%
  Adversarial Machine Learning, Sensitivity Analysis, Trajectory Forecasting, Autonomous Driving, Security of Cyber-Physical Systems%
\end{keywords}

\section{Introduction}
Trajectory prediction is becoming key for autonomy.  However, the integration of learning-based trajectory prediction into autonomous systems introduces new opportunities for cyber-attacks [\cite{miller2018securing}]. While there are efforts for integrating safety in autonomous control [\cite{bahati2021multi, chou2021reachability, herbert2021scalable}], inaccurate predictors of other agents' behaviors can adversely affect the navigation of an autonomous vehicle. Furthermore, even though recent work has shown several examples of possible perturbation attacks on trajectory predictors' state histories [\cite{cao2023robust, cao2022advdo, tan2023targeted, zhang2022adversarial}], because many of these models are multi-modal (taking in different input representations), there is still an open question as to whether or not perturbations on other input types, such as image maps, can adversely affect the performance of trajectory predictors. 

Therefore, we conduct a sensitivity analysis of both Trajectron++ [\cite{salzmann2020trajectron++}] and AgentFormer [\cite{yuan2021agentformer}] to understand the perturbation sensitivities of each model.  We then analyze how specially crafted perturbations, when applied to a vehicle using Trajectron++, produce erroneous predictions and change the resulting plans of a vehicle. The analyses show that both Trajectron++ and AgentFormer are most sensitive to perturbations on the most recent position and velocity states, demonstrating that effective attacks on both models can be made to just the current state of the input data. However, more interestingly, we observe that both models are susceptible to image attacks, showing that very small image perturbations can increase model error by at least $17-200 \%$ in either model. If it is given a well-designed perturbation informed by the model's sensitivity profile, a vehicle using an optimization-based planner with one of these models can suddenly come to a stop while traveling at speeds of around $\currSpeedRes$ (m/s). 

Our contributions are as follows: 
\begin{enumerate}
    \itemsep=0em
    \item A sensitivity analysis for security of trajectory prediction models that preserves outlier information; \label{Contr: sens_anlys}
    \item An illustration of the effects of image perturbation attacks on multi-modal trajectory prediction models; \label{Contr: image_pert}
    \item An analysis of how adversarial perturbations to a prediction model, crafted from the sensitivity results, can affect downstream vehicle planning. \label{Contr: planning}
\end{enumerate}

We conduct the sensitivity analysis at the level of each model's API, which, in this context, means that each perturbation is passed as an input to the model's main inference, or prediction, function. Along with the perturbations we consider, this analysis design considers attacks affecting data integrity that a user can make, without internal access to the model. Such attacks include ``Vehicle-to-everything (V2X)'' attacks, or attacks on vehicle sensors. 

We organize this paper as follows.  Background about the prediction problem and the two predictors under study are in Section \ref{Sec: Background}, and related works are provided in Section \ref{sec: Related Works}.  In Section \ref{Sec: Method}, we present the sensitivity analysis framework, then we provide the sensitivity results and discuss image perturbations in Section \ref{Sec: Results}. A demonstration of these results' impact on planning is provided in Section \ref{sec: Planning}.  Finally, security and learning implications are given in Section \ref{Sec: Implications}. Throughout the paper, the $\hat{}$ symbol denotes a predicted variable; for instance, $\predstate$ denotes a predicted state.

\section{Background: The Trajectory Prediction Problem, Trajectron++, and AgentFormer}
\label{Sec: Background}
Consider a system of $n$ agents. Let $\trajstate^\agentidx_\timeidx \in \mathbb{R}_{D}$ denote the state of vehicle $\agentidx$ at time $\timeidx$, which could include its position, velocity, heading, and angular acceleration. Let the time series of an agent's state $\trajstate^\agentidx_{0:\horizon} = (\trajstate^\agentidx_0, ..., \trajstate^\agentidx_\horizon)$ be known as the trajectory of that agent from time 0 to $\horizon$, and $\contextualInfo_\timeidx$ represent contextual information used for prediction (i.e an image map, $\contextualInfo_\timeidx \in \mathbf{R}^{W \times H \times L}$), where $W$, $H$, and $L$ are the width, height and number of channels in the image respectively. The trajectory prediction problem is to design a prediction model $\predictionModel(\trajstate^{[1:n]}_{\timeidx-\historyhorizon:\timeidx}, \contextualInfo_\timeidx, \cdot) \rightarrow \predstate^{[1:n]}_{\timeidx+1:\timeidx+\predictionhorizon}$ that predicts a future trajectory for agent $\agentidx$, $\predstate^\agentidx_{\timeidx+1:\timeidx+\predictionhorizon}$, from $\timeidx+1$ to $\timeidx+\predictionhorizon$ given the state history information of that agent $\trajstate^i_{\timeidx-\historyhorizon:\timeidx}$, state information of the other agents in the scene $\trajstate^{[1:n] \backslash i}_{\timeidx-\historyhorizon:\timeidx}$, and other contextual information accessible to the predictor. $\predictionhorizon$ and $\historyhorizon$ are the prediction horizon and number of previous time steps respectively.

To solve the trajectory prediction problem, both Trajectron++ and AgentFormer utilize a latent variable model to encode the distribution of future trajectories, in which the distribution of a value of interest, $\latentPredVar$, is represented as $p(\latentPredVar | \latentGiven) = \int p_{\latentPredDist}(\latentPredVar | \latentGiven, \latentVar) p_{\latentDist}(\latentVar | \latentGiven) \,d\latentVar$, where $\latentPredDist$ and $\latentDist$ are parameters of their respective distributions. Each trajectory predictor uses an encoder, $\encoder(\trajstate^{i}_{\timeidx-\historyhorizon:\timeidx}, \contextualInfo_\timeidx, \cdot) \rightarrow \latentDist$ to predict the distribution over the latent variables $\latentDist$ given prediction model inputs, and a decoder, $\decoder(\latentVar, \cdot) \rightarrow \latentPredDist$, to output the distribution $\latentPredDist$ over future agent actions ($\latentPredVar = \agentaction_{\timeidx+1:\timeidx+\predictionhorizon}$ for Trajectron++) or future agent positions ($\latentPredVar = \trajstate^\agentidx_{\timeidx+1:\timeidx+\predictionhorizon}$ for AgentFormer), where $\latentPredDist = (\mu, \sigma)$ represents Gaussian distributions. Both models use position, velocity, and heading as states and image maps (high-definition maps) for optional contextual information. Trajectron++ uses a spatio-temporal graph (``scene graph'') of the prediction scene as an extra input, $\graph = (\nodeSet, \edgeSet, \weightSet)$, encodes inputs using a series of Long Short-Term Memory (LSTMs) models and a Convolutional Neural Network (CNN), and passes either a sample or the mode of $\latentPredDist$ (depending on the user's specification) to a dynamics model $\agentdynamicsmodel: \agentaction^\agentidx_\timeidx, \predstate^\agentidx_\timeidx \rightarrow \predstate^\agentidx_{\timeidx+1}$ to produce the state predictions. Nodes of the scene graph, $\agentnode_j \in \nodeSet$, represent agents in the scene, edges ($\agentnode_i$, $\agentnode_j$) $\in \edgeSet$ represent the interaction between $\agentnode_i$ and $\agentnode_j$, and the weights $\weightSet_i \in \weightSet$, associated with each directed edge $i$, represent the amount of influence that agent $\agentnode_i$ has on agent $\agentnode_j$. AgentFormer is an augmented socio-temporal transformer model [\cite{vaswani2017attention}] that encodes inputs using a specialized attention module (treating $\trajstate^{[1:n] \backslash i}_{\timeidx}$ differently than $\trajstate^{\agentidx}_{\timeidx}$) and includes a multilayer perceptron (MLP) to compute the latent variable.  AgentFormer decodes trajectory states $\trajstate^{i}_{\timeidx}$ concatenated with the latent variable with a cross-attention mechanism (parameterized by the encoding) and an MLP. 

% High-definition maps are represented as $\contextualInfo_t \in \mathbf{R}^{W, H, L}$, where each layer $1 \leq l \leq L$ contains either geometric information or information about features of the map such as road boundaries, lane markings, and crosswalks.

% The image maps used by these trajectory predictors can range in fidelity from being simply obstacle maps, $\contextualInfo_t \in {0,1}^{W, H, 1}$, to HD maps, $\contextualInfo_t \in \mathbf{R}^{W, H, L}$, where each layer $1 \leq l \leq L$ contains either geometric information or information about features of the map such as road boundaries, lane markings, and crosswalks.

\section{Previous Approaches to Model Sensitivity Analysis and Feature Attribution}
\label{sec: Related Works}
Sensitivity analysis has been used to understand model sensitivities on image features and their contributions toward model output. Some methods include using partial derivatives as a measure [\cite{novak2018sensitivity, montano2003numeric, gevrey2003review, gevrey2006two}], measuring changes in the model's output response as each input dimension is perturbed [\cite{gevrey2003review}], using aggregated product of hidden neuron weights as a measure [\cite{montano2003numeric}], tracking changes in inner network activations [\cite{novak2018sensitivity}], and changing network configurations and their input representations [\cite{zhang2015sensitivity}]. Each method has its trade-offs - for example, while using partial derivatives can provide a lot of information about a particular input feature’s relation to the model, using just raw gradient values alone may not be optimal [\cite{smilkov2017smoothgrad, sundararajan2017axiomatic}] as features that are actually important may have small local gradients. Other methods of explaining feature importance are further explored in feature attribution literature. Methods either take a surrogate approach by analyzing feature importance on an equivalent but more interpretable network [\cite{ribeiro2016should}], use forward-propagated or back-propagated gradient values [\cite{simonyan2013deep, selvaraju2017grad, binder2016layer, shrikumar2017learning}], directly analyze connections between layer activations and input features [\cite{zeiler2014visualizing}], or take a game-theoretic approach by making use of Shapley values [\cite{sundararajan2020many, lundberg2017unified, makansi2021you}]. In contrast to these methods, this work attempts to contribute features across input modalities (i.e. image and non-image features) to model perturbation sensitivities on trajectory predictors for security and empirically observe the effects of these perturbations on vehicle control.

\section{Attributing Model Perturbation Sensitivities to Features}
\label{Sec: Method}
In our sensitivity analysis, we wish to quantify how much a specific perturbation on each input (referred to as input features) contributes to a change in the performance of $\predictionModel(\cdot)$. Therefore, we design a sensitivity attribution function, $\attrfunction(\predictionInput_1,...,\predictionInput_\numfeatures, \pertfunc(\cdot), \predictionModel(\cdot)): \rightarrow \attrscore_1,...,\attrscore_\numfeatures$, such that, given $\numfeatures$ number of input features, ($\predictionInput_1,...,\predictionInput_\numfeatures$), and a specific perturbation function, $\pertfunc(\predictionInput_i) \rightarrow \pertpredictioninput_{i}$, the function assigns scores, $\attrscore_\featureidx \in \mathbf{R}$, to each feature $\featureidx$, where each score indicates how much the perturbed input $\pertpredictioninput_{i}$ contributes to the observed change in the model's performance behavior. We use a common metric to measure model performance - average displacement error (see Definition \ref{def:Average Displacement Error}) - and choose the performance behavior for attribution to be the percent increase in average displacement error (Definition \ref{def:Percent Increase}). Given a dataset $\dataset_\datasetidx$ and a set of perturbation functions $(\pertfunc_1,...,\pertfunc_\numpertfuncs)$, we apply attribution function $\attrfunction(\cdot)$ to each data point $\datapoint \in \dataset_\datasetidx$, using each perturbation function $\pertfunc_\pertfuncidx(\cdot)$, and we evaluate the sensitivity behavior of the model $\predictionModel(\cdot)$ over $\dataset_{\datasetidx}$ for each function. Each perturbation function is described in Section \ref{Sec: Choosing_pert_func}. Results are aggregated into sets $\resultSet_{\pertfunc_\pertfuncidx, \dataset_\datasetidx} := \{\attrscore_i, i\in [1...|\dataset|] \}$ and then modeled as distributions, where each set corresponds to a specific perturbation function $\pertfunc(\cdot)$ and contains the sensitivity measures from applying $\attrfunction(\cdot)$ to each $\datapoint \in \dataset_\datasetidx$ using $\pertfunc(\cdot)$. To compute the feature to which the model is most sensitive, we compare quartile values of aggregated distributions using the conditions stated in Defintion \ref{def:sens}. Because it is possible for distributions across features to be similar, if no feature satisfies the condition in Definition \ref{def:sens}, then we use qualitative measures to draw conclusions, measuring distribution peaks, skews, and spreads.

\begin{definition}[Average Displacement Error]
    The average displacement error of a prediction $\predstate_{t+1:t+\predictionhorizon}$ over a prediction horizon of $\predictionhorizon$ is defined as 
    \begin{align}
        \text{ADE}(\predstate_{t+1:t+\predictionhorizon}, \trajstate_{t+1:t+\predictionhorizon}) = \frac{1}{\predictionhorizon}\sum_{i=t+1}^{t+\predictionhorizon} \|\predstate_{i, \text{pos}} - \trajstate_{i, \text{pos}} \|_2
    \end{align}
    where $\predstate_{i, \text{pos}}$ and $\trajstate_{i, \text{pos}}$ represents the prediction and ground truth position states at time $i$.
\label{def:Average Displacement Error} 
\end{definition}

\begin{definition}[Percent Increase as Sensitivity Measure]
    Let $\predstate_{t+1:t+\predictionhorizon} = \predictionModel(\predictionInput_1,...,\predictionInput_\numfeatures)$ be the baseline trajectory prediction from model $\predictionModel(\cdot)$ and $\predstate^{'}_{t+1:t+\predictionhorizon} = \predictionModel(\predictionInput_1,...,\pertfunc(\predictionInput_i), ...,\predictionInput_\numfeatures)$ be the new prediction resulting from applying perturbation function $\pertfunc(\cdot)$ on feature $\predictionInput_i$. The percent increase $\attrscore_i$ measured by the sensitivity function is defined as follows:
    \begin{align}
        \attrscore_i = \frac{\text{ADE}(\predstate^{'}_{t+1:t+\predictionhorizon}, \trajstate_{t+1:t+\predictionhorizon}) - \text{ADE}(\predstate_{t+1:t+\predictionhorizon}, \trajstate_{t+1:t+\predictionhorizon})}{\text{ADE}(\predstate_{t+1:t+\predictionhorizon}, \trajstate_{t+1:t+\predictionhorizon})}
    \end{align}
    If $\text{ADE}(\predstate_{t+1:t+\predictionhorizon}, \trajstate_{t+1:t+\predictionhorizon}) = 0$, we aggregate this data separately.
\label{def:Percent Increase} 
\end{definition}
 
\begin{definition}[Condition for Feature with Most Model Sensitivity]
    Let the quartile metrics $[\qone_\resultSetidx, \qtwo_\resultSetidx, \qthree_\resultSetidx]$ of set $\resultSet_\resultSetidx$ represent first quartile, median, and third quartile values respectively. Feature $\predictionInput_\featureidx$ with quartile measures $[\qone_\featureidx, \qtwo_\featureidx, \qthree_\featureidx]$ has greatest model sensitivity if
    \begin{align}
    \qone_\featureidx > \qone_j, ~\qtwo_\featureidx > \qtwo_j, ~\qthree_\featureidx > \qthree_j ~~\forall j \in [1...\numfeatures] \backslash \featureidx
    \end{align}
\label{def:sens} 
\end{definition}

\subsection{Choosing Perturbation Functions and Input Features for Analysis}
\label{Sec: Choosing_pert_func}
Each network's architecture helps to inform what perturbation types we use for the study. If we understand each model's \textit{inductive bias}, we can understand what perturbations may or may not be effective. For example, CNN models have inductive biases that focuses on ``local features'' and, thus, will identify and process local patterns within an input image. As a consequence, a model using CNNs is invariant to certain input changes, such as input translation and rotation. Therefore, we can rule out a class of perturbations - for instance jittering and image rotations - and consider perturbations that change the local structure within an image, such as noise and occlusion perturbations. For this analysis, we choose to use noise, occlusions, gradient perturbations, Fast Gradient Sign Method (FGSM) [\cite{goodfellow2014explaining, kurakin2016adversarial}], and perturbations with constant values (all defined in Definition \ref{def:perts}). FGSM is among the perturbations chosen because it provides a method of generating perturbations that increases some cost associated with the model. For input features, we simply choose to perturb inputs accessible from eah model's API inference call that may contribute heavily to it's output prediction. These are: state histories $\trajstate^i_{\timeidx-\delta t_h:\timeidx}$ and image maps $\contextualInfo_\timeidx \in \mathbf{R}^{L \times W \times H}$, for both models, as well as scene graph components, $\nodeSet$ and $\weightSet$, for Trajectron++. 

\begin{definition}[Perturbation Types]
    Let $\predictionInput_i$ be the input feature in vector form $\predictionInput_i \in \mathbf{R}^{m}$. The perturbation function $\pertfunc(\cdot)$ produces the perturbed input $\pertpredictioninput_{i}$ as follows:
    \begin{align}
    \pertpredictioninput_{i} = \pertfunc(\predictionInput_i) = \predictionInput_i + \perturbation
    \end{align}
    where $\perturbation = \epsilon \nabla_{\predictionInput_i} J(\modelparams, \predictionInput, \inputtargets)$ for gradient perturbations, $\perturbation = \epsilon \text{sign}(\nabla_{\predictionInput_i} J(\modelparams, \predictionInput, \inputtargets))$ for a Fast Gradient Sign Method (FGSM) perturbation, $\perturbation = [c]_{1 \times m}$ for constant perturbations, $\perturbation_j \sim \mathcal{N}(0, \sigma)$ for noise, and $\perturbation = [0]_{1\times m}$ for occlusion. $J(\cdot)$ is the loss function of $\predictionModel(\cdot)$, $\modelparams$ are the parameters of $\predictionModel(\cdot)$, and $\inputtargets$ are the ground truth states $\trajstate^\agentidx_{\timeidx+1:\timeidx+\predictionhorizon}$ associated with the unperturbed baseline predictions $\predstate^\agentidx_{\timeidx+1:\timeidx+\predictionhorizon}$. $\epsilon$, $c$, and $\sigma$ are parameters chosen based on the normalization applied (described in Section \ref{sec:init_Tplus}).
\label{def:perts} 
\end{definition}

% \subsection{Threat Model}
% \label{Sec: threat_model}
% We conduct the sensitivity analysis at the level of each model's API, which, in this context, refers to the set of functions a user would call to make an inference on the model. Each perturbation is passed as an input at this level and predictions are collected from the return values of these functions; thus, simulating perturbation attacks that a user without internal access can make.

\section{Sensitivity Analysis Results: Trajectron++ and AgentFormer}
\label{Sec: Results}
In this section, we discuss the sensitivity analysis results for Trajectron++ and AgentFormer and highlight the effect of image perturbations. Learning and security implications are discussed in Section \ref{Sec: Implications}. Trajectron++ and AgentFormer are trained using the mini-NuScenes dataset [\cite{caesar2020nuscenes}] and data used for the analysis are from the training, validation, and testing partitions of the dataset. We use the same training procedures and loss functions as used in [\cite{salzmann2020trajectron++}] and [\cite{yuan2021agentformer}] respectively. Other possible data sets to evaluate/train on, which we leave for future work, include the full NuScenes dataset [\cite{caesar2020nuscenes}], ETH dataset [\cite{pellegrini2009you}], and UCY dataset [\cite{lerner2007crowds}]. 

\subsection{Contributing sensitivity scores to features: the effects of current state and image perturbations}
\label{sec:init_Tplus}

\begin{figure*}[t!]
    \centering
    \begin{subfigure}[t]
        \centering
        \includegraphics[scale=0.21]{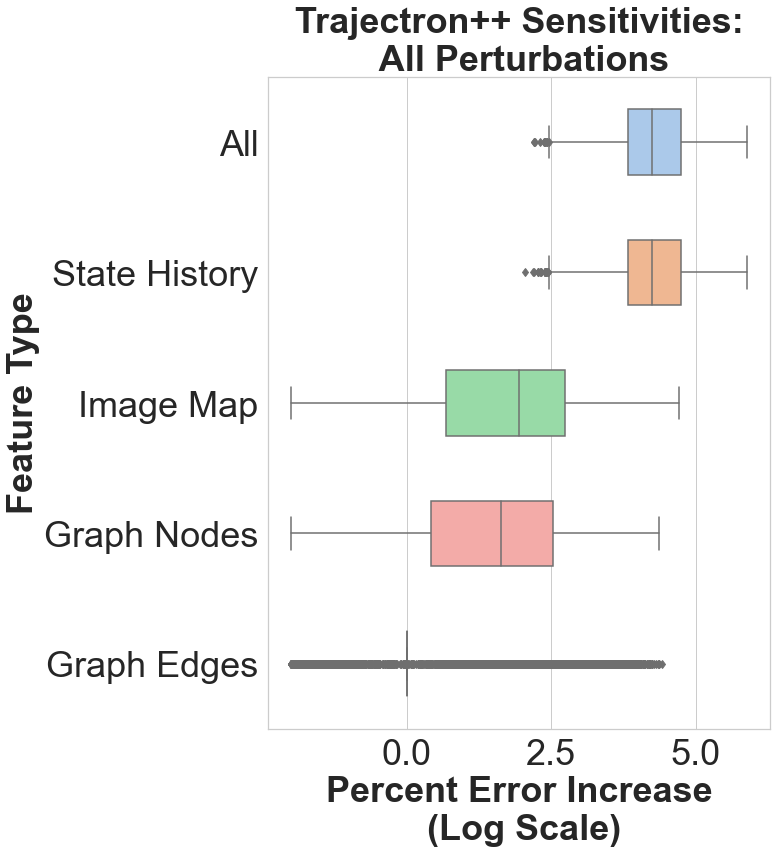}
        % \label{subfigure:nondepth_sens}
    \end{subfigure}
    \begin{subfigure}[t]
        \centering
        \includegraphics[scale=0.29]{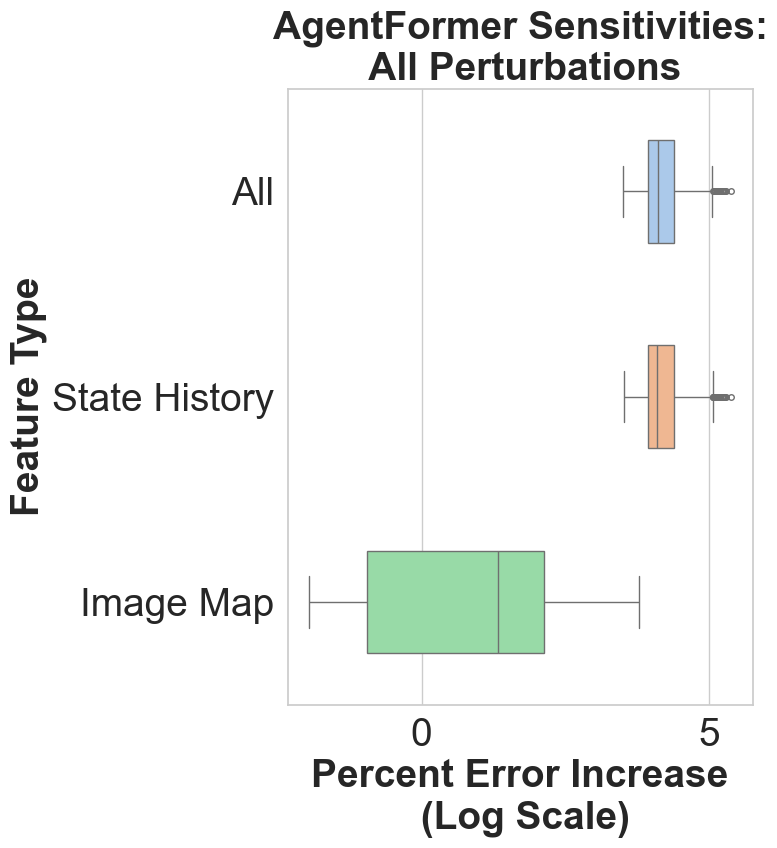}
        % \label{subfigure:FGSM_sens}
    \end{subfigure}
    \caption{\label{Fig: Sens_analysis} Sensitivity analysis of Trajectron++ (a) and AgentFormer (b) with power transformation [\cite{yeo2000new}] to values. Both models are by far most sensitive to perturbations on state history inputs - having mean sensitivity of (a) $17200 \%$ and (b) $12573.495 \%$ percent error increases. However, perturbations on image maps and graph nodes can induce at least a median of $20.5 - 85.5 \%$ and $ 41.8 \%$ increase in error respectively, indicating that perturbations on other inputs can adversely affect the performance of these trajectory predictors.  }
\end{figure*}

To observe characteristics of the models' sensitivities, we report the quartile values of aggregated results of two experiments - one for Trajectron++ and one for AgentFormer. Both experiments measures sensitivity as described in Definition \ref{def:Percent Increase}. For each experiment, each perturbation is normalized such that the magnitude of the perturbation made to a particular input is \normperc\% of that input's range in the dataset. All ranges are either taken from [\cite{caesar2020nuscenes}] or calculated from the dataset directly. We use a range of $\posrange \text{m}$,  $\velrange \frac{\text{m}}{\text{s}}$, $\accelrange \frac{\text{m}}{\text{s}^2}$, $\headrange \text{ rad}$, $\dheadrange \frac{\text{rad}}{\text{s}}$, $\imagerange \text{ units}$, and $\edgeweightrange \text{ units}$ for perturbations made on position, velocity, acceleration, heading, angular velocity, image, and edge weight values respectively. Ranges on standardized inputs are calculated similarly with respect to their standardized variables. The gradients used for gradient perturbations and FGSM are the gradients of a negative evidence lower bound (ELBO) loss function with respect to the baseline inputs. Perturbed image pixel values remain unclipped, and any baselines points with initial ADE values of 0 are left out of the aggregated data. To obtain a better view of the data's distribution, we apply a power transformation from [\cite{yeo2000new}] before plotting distributions in Figure \ref{Fig: Sens_analysis}. 

\begin{figure*}[t!]
\centering
\begin{subfigure}[t]
    \centering
    \includegraphics[scale=0.33]{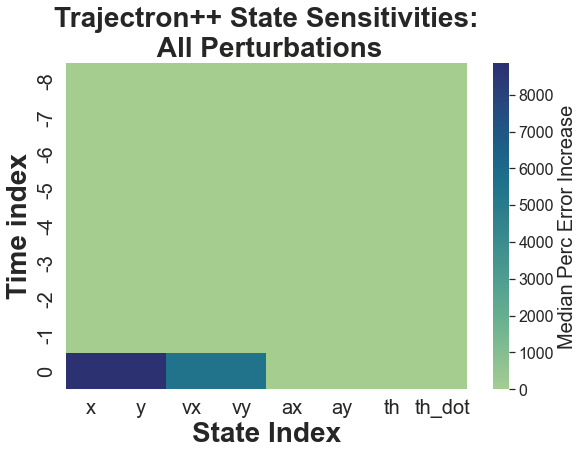}
    % \label{subfigure:nondepth_sens}
\end{subfigure}
\begin{subfigure}[t]
    \centering
    \includegraphics[scale=0.40]{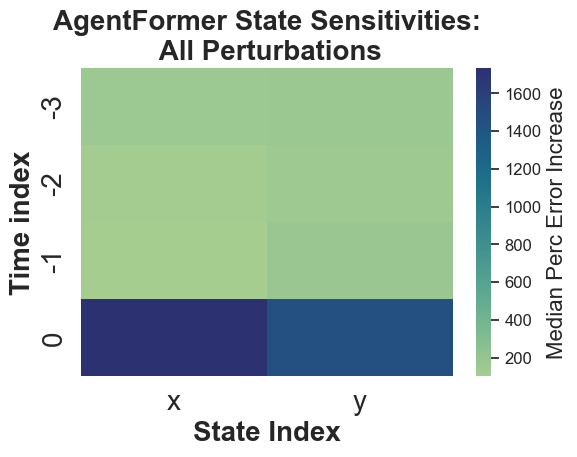}
    % \label{subfigure:FGSM_sens}
\end{subfigure}
\caption{Depth sensitivity analysis for Trajectron++ (a) and AgentFormer (b). We observe that all sensitivity is on the most recent position ($x$,$y$) and velocity states ($v_x$,$v_y$) for both models.}\label{Fig: Depth_Sens_analysis}
\end{figure*}

From the results, we observe that the quartile values for state history input are by far the largest and most similar to sensitivity measures from perturbing all inputs on both Trajectron++ and AgentFormer. Trajectron++'s median perturbation sensitivity on state history inputs is two to four hundred times as much as on image and graph node inputs respectively, while AgentFormer's is more than six hundred times as much as on image inputs. Given that results show high perturbation sensitivity on state history features, we run a ``depth analysis'' to study perturbation sensitivities on each state at different time points, shown in Figure \ref{Fig: Depth_Sens_analysis} (no data transformation applied). We observe, for both models, that most of the sensitivity contribution for state history inputs comes from the most recent position and velocity states, where all other values have median sensitivity measures of less than $\Tppminordepthres \%$ for Trajectron++ and less than $\AgFminordepthres \%$ for AgentFormer. This observed phenomenon is due to the implementation and architectural choices used in both models. For Trajectron++ and AgentFormer, the current position and velocity states are used as either the initial condition for Trajectron++'s dynamics model (see Section \ref{Sec: Background}) or used to translate AgentFormer's predictions from relative coordinates back to absolute coordinates. Therefore, any perturbation to the initial condition or offset will drastically perturb the resulting predictions. For instance, if we keep the initial condition for Trajectron++'s dynamic model unperturbed, we see that Trajectron++ becomes most sensitive to image perturbations, with median sensitivities on image features being $85.5 \%$ vs $83.4 \%$ for state history.

% \begin{figure*}[t!]
%     \centering
%     \begin{subfigure}[t]
%         \centering
%         \includegraphics[scale=0.23]{pics_version_2/Tpp_non_depth_no_offset.png}
%         % \label{subfigure:nondepth_sens}
%     \end{subfigure}
%     \begin{subfigure}[t]
%         \centering
%         \includegraphics[scale=0.33]{pics_version_2/AgF_non_depth_no_offset.png}
%         % \label{subfigure:FGSM_sens}
%     \end{subfigure}
%     \caption{\label{Fig: Sens_analysis_no_offset} Sensitivity analysis of Trajectron++ (a) and AgentFormer (b) without a perturbation to dynamic model initial condition or prediction offset (power transformation [\cite{yeo2000new}] applied to values. Here, we observe that Trajectron++ is now most sensitive to image perturbations while AgentFormer remains heavily sensitive to state history perturbations.}
% \end{figure*}

\begin{figure*}[t!]
    \centering
    \begin{subfigure}[t]
        \centering
        \includegraphics[scale=0.22]{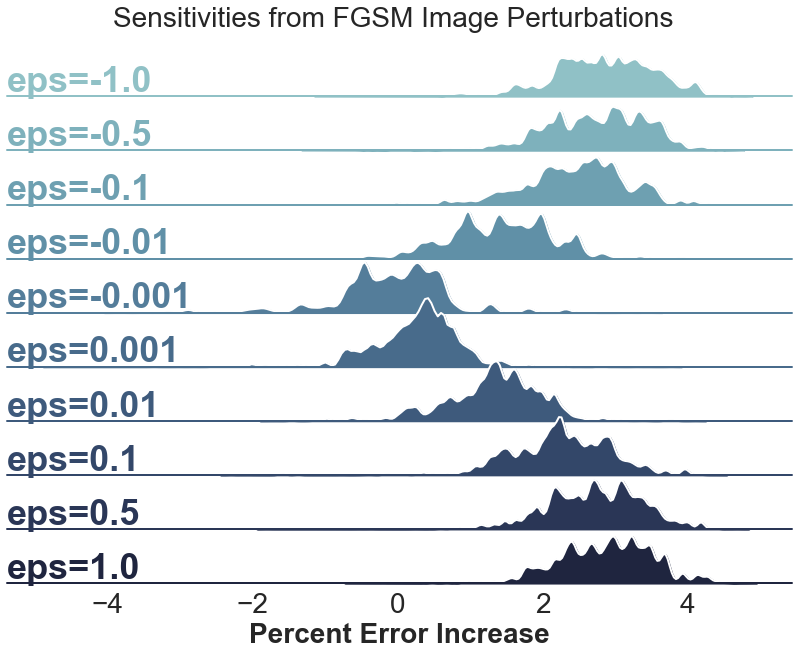}
        % \label{subfigure:nondepth_sens}
    \end{subfigure}
    \begin{subfigure}[t]
        \centering
        \includegraphics[scale=0.22]{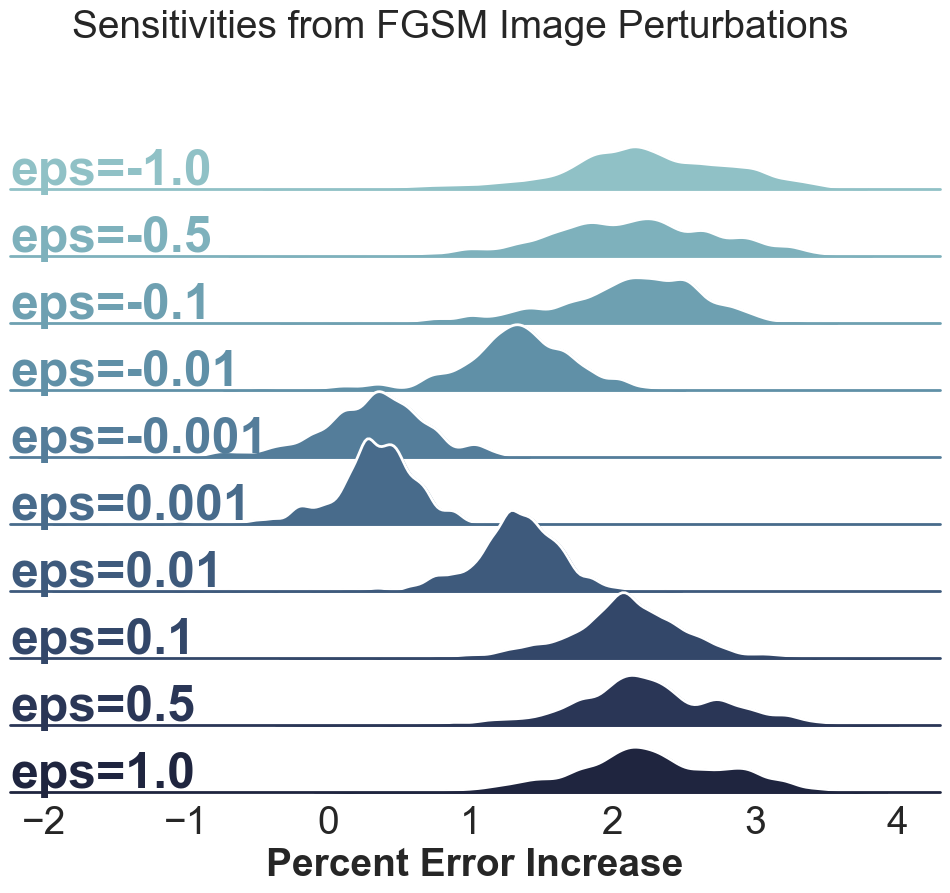}
        % \label{subfigure:FGSM_sens}
    \end{subfigure}
    \caption{\label{Fig:FGSM_Image_analysis} Sensitivity analysis of image perturbations for Trajectron++ (a) and AgentFormer (b) with power transformation [\cite{yeo2000new}] applied to values. Percent error increases from image perturbations with epsilon size as small as 0.01 contain median values of $\TppZZOneMedian \%$ and $\AgFZZOneMedian \%$ and go as large as $\TppZZOneMax \%$ and $\AgFZZOneMax \%$ for Trajectron++ and AgentFormer respectively.}
\end{figure*}

\begin{figure*}[t!]
    \centering
    \begin{subfigure}[t]
        \centering
        \includegraphics[scale=0.15]{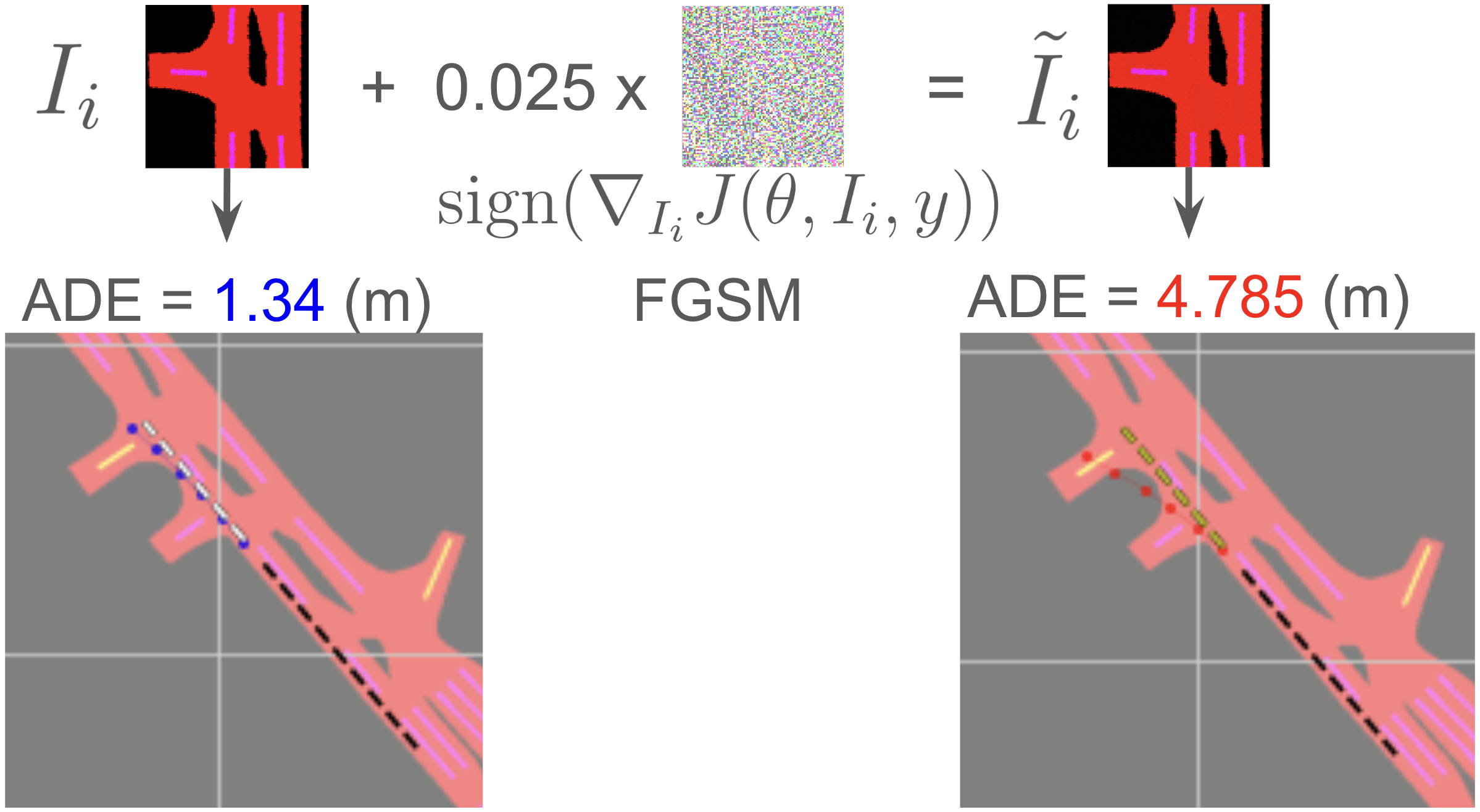}
        % \label{subfigure:nondepth_sens}
    \end{subfigure}
    \begin{subfigure}[t]
        \centering
        \includegraphics[scale=0.15]{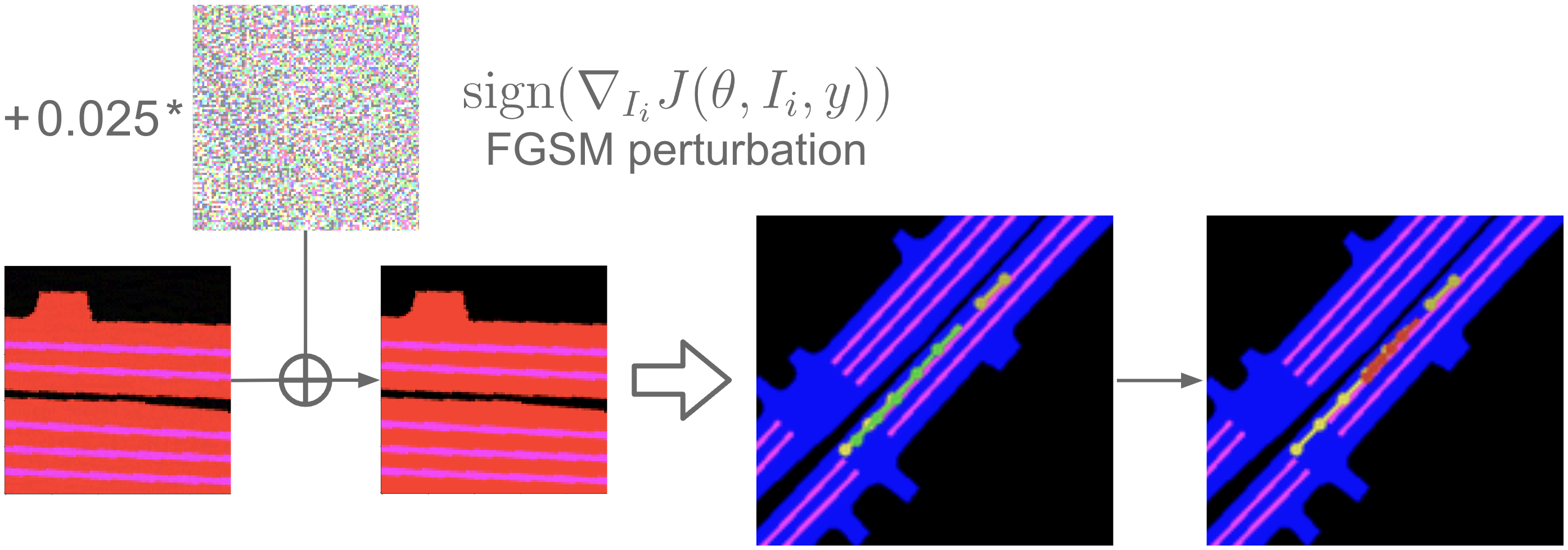}
        % \label{subfigure:FGSM_sens}
    \end{subfigure}
    \caption{\label{Fig: im_pert_ex} Image perturbation examples for Trajectron++ (a) and AgentFormer (b). These undetectable image perturbations cause a (a) $\undetectRes \%$ and (b) $\AFundetectRes$ \% error increase in Trajectron's and AgentFormer's ADE respectively, where (b) goes from $\AFADEFrom$ to $\AFADETo$ (m) ADE. Here we show that, with the inclusion of image inputs, these multi-modal generative models are susceptible to image-based attacks.}
\end{figure*}

Because both Trajectron++ and AgentFormer have an upper quartile image perturbation sensitivity of $\TppUpperQuartileImage \%$ and $\AgFUpperQuartileImage \%$ respectively, we run an additional experiment to observe how different sized image perturbations affect the models' prediction errors. The results, shown in Figure \ref{Fig:FGSM_Image_analysis}, indicate that an image perturbation size, as small as $0.01$ units per dimension, can cause a median of $\TppZZOneMedian \%$ and $\AgFZZOneMedian \%$ increase in error for Trajectron++\footnote{Because Trajectron++ outputs a Gaussian Mixture Model, we analyze whether there is mode switching in the output as the size of perturbations increases on images. We observe that frequent mode switching seems to occur for image perturbations, when increasing perturbation sizes steadily between $0$ and $1$. We compare this to observations of mode switching on state histories inputs, which is far less frequent as perturbation sizes increase between $0-\posrange \text{m}$.} and AgentFormer respectively, while a perturbation size of $0.1$ can increase error on both models by $\TppZOneMedian \%$ and $\AgFZOneMedian \%$. Therefore, we demonstrate examples of image perturbations with sizes in between both values that have significant impact (within the upper quartile of sensitivity) on both Trajectron++ and AgentFormer. As shown in Figure \ref{Fig: im_pert_ex}, one undetectable image perturbation with a size $\tplusFGSMEp$ per dimension alone can cause Trajectron++'s and AgentFormer's prediction error to go from $\tplusADEFrom$ (m) to $\tplusADETo$ (m) and  $\AFADEFrom$ (m) to $\AFADETo$ (m). 

% \begin{figure*}[t!]
%     \centering
%     \begin{subfigure}[t]
%         \centering
%         \includegraphics[scale=0.15]{pics/image_pert_ex.png}
%     \end{subfigure}
%     \caption{\label{Fig: im_pert_ex} Trajectron++ image perturbation example. This undetectable image perturbation causes a \undetectRes \%  increase in the model's ADE. Here we show that, as demonstrated on image classifiers in \cite{goodfellow2014explaining}, this same effect hold for this multi-modal generative model suggesting that these models, with the inclusion of image inputs, can become susceptible to the same type of image attacks.}
% \end{figure*}

% \begin{figure*}[t!]
%     \centering
%     \begin{subfigure}[t]
%         \centering
%         \includegraphics[scale=0.2]{pics/AF_Image_Example.png}
%     \end{subfigure}
%     \caption{\label{Fig: AF_im_pert_ex} AgentFormer image perturbation example. This undetectable image perturbation, similar to Trajectron++, causes a $\AFundetectRes$ \% error increase in this model's ADE. $\AFADEFrom$ to $\AFADETo$ (m).}
% \end{figure*}

\subsection{Discussion}
Given these observations, it is clear that perturbations made to state history features would be most effective for attacking both models; however, it is unclear as to what perturbation sensitivities are directly learned in training or can be attributed to neural network components alone. We leave this investigation for future work. Regarding the phenomenon observed in Figure \ref{Fig: im_pert_ex}, our hypothesis is that, if an input's dimensions are large enough as in the case with image maps, even small perturbations per dimension, if well designed, can be enough to cause some significant change in model performance. 

\section{Downstream Effects on Planning}
\label{sec: Planning}
Next, we demonstrate how adversarial perturbations to trajectory prediction models could affect downstream planning. In this section, we take an example image and state history perturbation, informed by the sensitivity analysis, and observe changes in an example planner using Trajectron++.

\subsection{Scenario and Planner Analysis}
\begin{figure*}[t!]
    \centering
    \begin{subfigure}[t]
        \centering
        \includegraphics[scale=0.4]{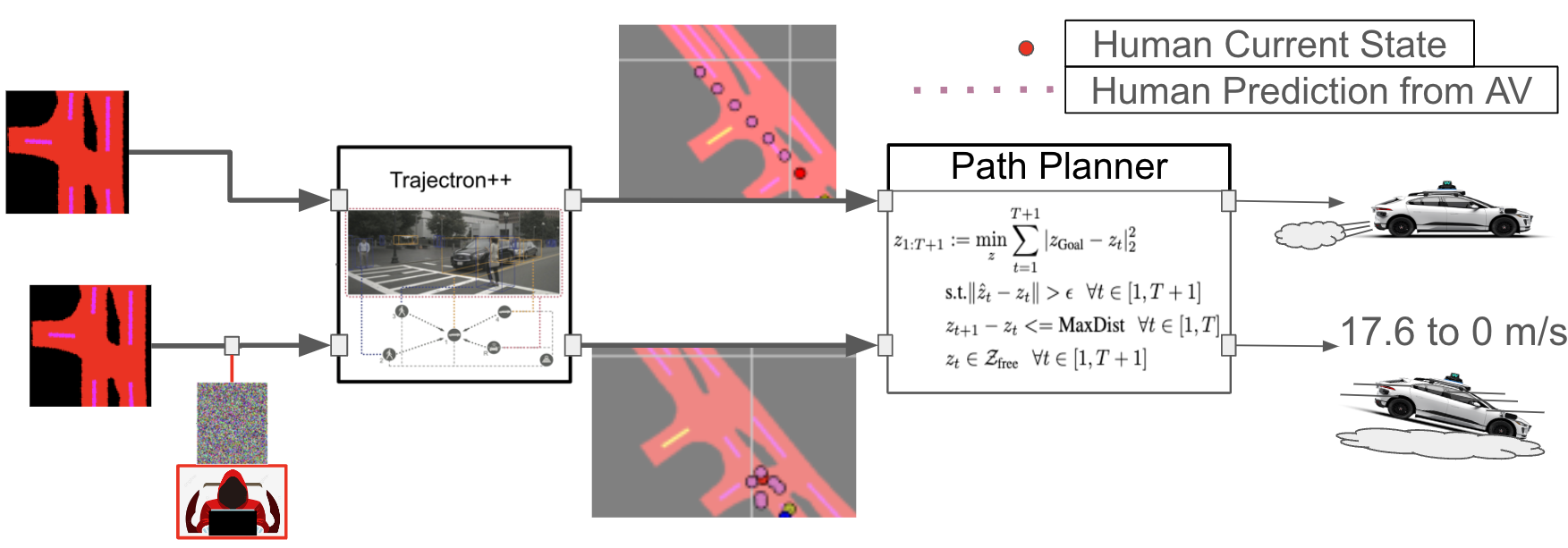}
    \end{subfigure}
    \caption{\label{Fig: image_onplanner} Example effect of image perturbation on vehicle planning. Top path shows the baseline input, where the autonomous vehicle (AV) w/ Trajectron++ produces a ``driving forward'' prediction for the human vehicle, allowing it to continue driving at same previous speed. The bottom path shows the prediction and plan with the perturbation, which causes the AV’s human vehicle prediction to clump together, causing it to come to an abrupt stop.}
\end{figure*}

To analyze effects of perturbations on vehicle planning, we take a real driving scenario from the mini-NuScenes dataset that contains at least two ``interacting'' vehicles in close proximity to each other (where the behavior of one vehicle impacts another) and see how the model's input perturbations, chosen based on sensitivity analysis, are propagated through to an optimization-based planner. We chose a two-vehicle scenario, one following the other, where the following vehicle is chosen to be the autonomous vehicle (AV) predicting with Trajectron++ and planning with the generic planner, and the leading vehicle is chosen to be the human vehicle whose predicted behavior impacts the future plans of the AV. Our optimization-based planner, which captures the basics behind many different planners, is:
\begin{align}
    \trajstate_{1:T+1}& := \min_{\trajstate} \sum_{t=1}^{T+1} |\trajstate_{\textrm{Goal}} - \trajstate_t|_{2}^2 \\
    &\textrm{s.t.} \| \predstate_t - \trajstate_t \|_2 > \validHumanDist
    ~~\forall t \in [1, T+1] \\
    & | (\trajstate_{t+1})_i - (\trajstate_{t})_i| <= \validplanDist ~~\forall i \in [1,\numPosStates], t \in [1,T] \\
    & \trajstate_{t} \in \mapfree ~~\forall t \in [1,T+1]\
\end{align}
Given the goal position $\trajstate_{\textrm{Goal}}$, and predictions $\predstate_t$ from the trajectory predictor $\text{Trajectron}(\cdot)$, the planner produces planned states $\trajstate_{1:T+1}$. Planned states must be on valid road space $\mapfree$, the distance between each subsequent planned state must be less than or equal to $\validplanDist$, and each $\trajstate_t$ should be at least $\validHumanDist$ units away from the predicted human vehicle state $\predstate_t$. The parameters $\validHumanDist=\humanDistNum$ and $\validplanDist=\planDistNum$ are chosen such that the resulting planned states approximates the vehicle's (which represents the AV) real states in the NuScenes scenario. Figure \ref{Fig: image_onplanner} highlights an example of the effects that an image perturbation has on the planning of the vehicle utilizing Trajectron++. This perturbation, made with FGSM using epsilon size of $\planFGSMep$, causes Trajectron++ to output an erroneous prediction for the human vehicle (jumping from an ADE of $\planPrevADE$(m) to $\planToADE$(m)), which causes the autonomous vehicle's future plan to induce an abrupt stop, coming from speeds of around $\planPrevSpeed$(m/s) to $\planPertSpeed$. The same effect happens when the most recent velocity states get set to 0, see Table \ref{table: planner_analysis}. If driving on a freeway or major street, these perturbations could cause this autonomous vehicle to collide with any vehicle following behind them. 

% The same effect happens when the most recent velocity states are perturbed. Table \ref{table: planner_analysis} shows that when the most recent velocity states get set to 0, Trajectron++ produces a stop prediction causing the autonomous vehicle to similarly stop abruptly. 

\begin{table}[!ht]
\begin{center}
\resizebox{0.75\columnwidth}{!}{%
\begin{tabular}{|c|c|c|c|c|c|}
\hline
 & \textbf{t=0} & \textbf{t=1} & \textbf{t=2}  & \textbf{t=3}  & \textbf{t=4} \\
\hline
\textbf{Baseline Future States} 
& \textbf{x}: 619.23 & 609 & 593 & 577 & 561 \\
& \textbf{y}: 482.87 & 499 & 515 & 531 & 547 \\
\hline
\textbf{Planning after Image Perturbation} 
& 619.23 & 619.23 & 619.23 & 619.23 & 619 \\
& 482.87 & 482.87 & 482.87 & 482.87 & 499 \\
\hline
\textbf{Planning after t=0 Velocity Perturbation} 
& 619.23 & 619.23 & 619.23 & 619.23 & 619.23 \\
& 482.87 & 482.87 & 482.87 & 482.87 & 482.87 \\
\hline
\end{tabular}
}
\caption{\label{table: planner_analysis} Results from the optimization-based planner. Both the image perturbation (made with FGSM) and occlusion on the most recent velocity state (both chosen based on Trajectron++'s sensitivity profile) causes erroneous predictions from Trajectron++, which consequently cause the autonomous vehicle’s future plans to simulate an abrupt stop.}
\end{center}
\end{table}

\section{Discussion: Implications on Learning and Security}
\label{Sec: Implications}
\textbf{Contribution \ref{Contr: sens_anlys}}. The sensitivity results show both models having dominant sensitivity on the most recent position/velocity states, and non-trivial sensitivity on image inputs. These results reveal that, although any targeted attack made to either Trajectron++ and AgentFormer can be done most effectively using the most recent position/velocity state input, other attacks can utilize non-state related inputs, such as image maps, to impact either model's performance.

\noindent \textbf{Contributions \ref{Contr: image_pert}}. The image perturbation analysis shows that an image perturbation with a size of $0.01$ units can induce a median of $\TppZZOneMedian \%$ and $\AgFZZOneMedian \%$ error increase in Trajectron++ and AgentFormer respectively, suggesting that both of these models are highly susceptible to image attacks, as shown in Figure \ref{Fig: im_pert_ex}. One possible mitigation may be to use techniques from \cite{ilyas2019adversarial}, which removes ``non-robust features'' from image data. 

\noindent \textbf{Contribution \ref{Contr: planning}}. We show that from the sensitivity analysis, we can craft targeted perturbations for Trajectron++ in the form of an image perturbation or an occlusion over the most recent velocity state, and that if these targeted perturbations are given as an input to a generic optimization-based planner utilizing Trajectron++, the planner will return a plan that stops the vehicle even if the vehicle is previously traveling at $\planPrevSpeed$ (m/s). When a trajectory prediction model is embedded in the autonomous vehicle's system, an adversarial example for the prediction model can become adversarial to the vehicle's planning and control, highlighting the significance of minimizing any neural network model's ``attack surface'' before integration into safety-critical systems for better robustness and security.

\section{Conclusion and Future Work}
We conducted a full sensitivity analysis on Trajectron++ and AgentFormer, illustrated the effect of image perturbations, and analyzed the impact of targeted image and state perturbations on downstream vehicle planning. We demonstrated that the effect of image perturbations exploits a possible trajectory model vulnerability that we highlighted in Section \ref{Sec: Implications}.  

\bibliography{main.bib}

\end{document}